\documentclass[pdflatex,sn-mathphys-num]{sn-jnl}% Math and Physical Sciences Numbered Reference Style
\usepackage{graphicx}%
\usepackage{multirow}%
\usepackage{amsmath,amssymb,amsfonts}%
\usepackage{amsthm}%
\usepackage{mathrsfs}%
\usepackage[title]{appendix}%
\usepackage{xcolor}%
\usepackage{textcomp}%
\usepackage{manyfoot}%
\usepackage{booktabs}%
\usepackage{algorithm}%
\usepackage{algorithmicx}%
\usepackage{algpseudocode}%
\usepackage{listings}%

\usepackage{soul}
\usepackage{tikz-cd}
\usepackage{comment}
\usepackage{amsmath}
\usepackage{bbm}
\usepackage{subcaption}
\usepackage{gensymb}
\theoremstyle{thmstyleone}%
\theoremstyle{thmstyletwo}%

\theoremstyle{thmstylethree}%

\begin{document}

\title[Estimating Inhomogeneous Spatio-Temporal Background Intensity Functions using Graphical Dirichlet Processes]{Estimating Inhomogeneous Spatio-Temporal Background Intensity Functions using Graphical Dirichlet Processes}


\author*[1]{\fnm{Isa\'ias} \sur{Ba\~nales}}\email{isaias@ciencias.unam.mx}

\author[1]{\fnm{Tomoaki} \sur{Nishikawa}}\email{nishikawa.tomoaki.2z@kyoto-u.ac.jp}
\equalcont{These authors contributed equally to this work.}

\author[1]{\fnm{Yoshihiro} \sur{Ito}}\email{ito.yoshihiro.4w@kyoto-u.ac.jp}
\equalcont{These authors contributed equally to this work.}

\author[2]{\fnm{Manuel J.} \sur{Aguilar-Vel\'azquez}}\email{manuel.aguilar.411@gmail.com}
\equalcont{These authors contributed equally to this work.}


\affil*[1]{\orgdiv{Disaster Prevention Research Institute}, \orgname{Kyoto University }, \orgaddress{\street{Gokasho}, \city{Uji}, \postcode{611-0011}, \state{Kyoto}, \country{Japan}}}

\affil[2]{\orgdiv{Department of Earth and Planetary Science}, \orgname{The University of Tokyo}, \orgaddress{\street{ 7 Chome-3-1 Hongo}, \city{Bunkyo}, \postcode{113-8654}, \state{Tokyo}, \country{Japan}}}


\abstract{ Enhancements in seismic measuring instrumentation have been proved to have implications in the quantity of observed earthquakes, since denser networks usually allow recording more events. However, phenomena such as strong earthquakes or even aseismic transients, as slow slip earthquakes, may alter the occurrence of earthquakes. In the field of seismology, it is a standard practice to model background seismicity as a Poisson process. Based on this idea, this work proposes a model that can incorporate the evolving spatial intensity of Poisson processes over time (i.e., we include temporal changes in the background seismicity when modeling). In recent years, novel methodologies have been developed for quantifying the uncertainty in the estimation of the background seismicity in homogeneous cases using Bayesian nonparametric techniques. This work proposes a novel methodology based on graphical Dirichlet processes for incorporating spatial and temporal inhomogeneities in background seismicity. The proposed model in this work is applied to study the seismicity in southern Mexico, using recorded data from 2000 to 2015.}




\maketitle

\section{Introduction}
\label{sec:Intro}
Since its introduction, the Epidemic Type Aftershock Sequence (ETAS) model \citep{ogata1988statistical} has proved to be one of the most important tools for assessing stochastic seismicity, in which earthquake activity is modeled using Hawkes process \citep{Hawkes1971Spectra}. In this context, the work by \citep{ogata1988statistical} assumed that the mainshocks (or background earthquakes) originate from a Poisson process.

Recently, due to the importance of having uncertainty quantification of the estimators, authors such as \cite{ross2022semiparametric} and \cite{molkenthin2022gp} have worked with Bayesian nonparametric approaches to model a temporally homogeneous background intensity function $\mu(x,y)$ in equation \eqref{eq:IntensityHawkes} using Dirichlet processes mixtures (DPM) \citep{antoniak1974mixtures} and Gaussian processes, respectively.

Since the homogeneity of the ETAS model can be affected by tectonic or pore pressure changes \citep{hainzl2005detecting,hainzl2013impact,marsan2013monitoring}, such phenomena may have implications for the occurrence of tremors. However, these changes can be produced by strong earthquakes \cite{cruz2021short}, and can also lead to an increase in the seismicity rate \citep{nishikawa2017detection,banales2025inhomogeneous}.

Furthermore, the presence of slow slip events—slow fault slip phenomena that mainly occur along the plate interface—can alter seismic activity \citep{marsan2013slow,fukuda2018variability,nishikawa2023development}, as well as periods of microearthquake quiescence \citep{matthews1988statistical}. Moreover, not only variations in seismicity can affect ETAS model estimates, but also factors such as short-term aftershock incompleteness \citep{hainzl2016apparent,asayesh2025improved} and changes in the cut-off magnitude \citep{hainzl2022etas}. These temporary changes in activity emphasize the need for inhomogeneous models over time for the background seismicity function, such as the one introduced in this paper.

% ------------------------------------------------------
Consequently, this work contribution lies on the following key points:

\begin{enumerate}
    \item We propose a nonparametric Bayesian approach to estimate the background seismicity function based on Graphical Dirichlet Processes (GDP) \citep{chakrabarti2024graphical}, which allows for inhomogeneities in both space and time.
    \item We focus on inferring the background intensity adding inhomogeneities in time, i.e. $\mu(x,y,t)$, following a GDP as proposed by \cite{chakrabarti2024graphical}. In the context of tectonic activity, this variation of the background seismic activity model holds significance, as it may provide valuable insights into the actual tectonic physics that govern earthquake occurrence.
    \item We only study the background intensity function because the \cite{sawires2019updated} database is used. Its main feature lies in the fact that aftershocks, foreshocks, and earthquake swarms were already removed using the methodology proposed by \cite{gardner1974sequence}, leaving only mainshocks (background events) available (i.e., we will only consider the Poissonian intensity).
\end{enumerate}

In the seismological literature, there are two main approaches to distinguish background events from aftershocks: those based on deterministic spatio-temporal filtering, as proposed in \cite{gardner1974sequence} and used in \citep{sawires2019updated}, and those based on stochastic models following the ETAS model proposed in \citep{ogata1988likelihood}, as in \cite{zhuang2002stochastic,schoenberg2013facilitated}. It is important to emphasize that earthquake modeling corresponds to a self-exciting point process, and while, as previously mentioned above in point 3, we will use the catalog of \cite{sawires2019updated}, any assumption about the background intensity function may lead to potential biases in the estimation of the triggering effects.

While the catalog of \citep{sawires2019updated} allows us to focus only on the Poissonian part of the problem, in Section \ref{sec:Methodology} we describe how our work can generate extensions of the model proposed in \citep{ross2022semiparametric}, based on the ETAS model, by incorporating hidden variables in a manner similar to that developed in \citep{schoenberg2013facilitated,banales2025inhomogeneous}.

% ------------------------------------------------------

In the frequentist context, \cite{li2020space} has used piecewise constant functions and kernel density estimators to model an inhomogeneous space-time background activity function in the context of the Hawkes process. They assumed that
$$\mu(x,y,t)=\gamma f(x,y)v(t),$$
where $\gamma\in \mathbb{R}$, $f$ is a probability density function, and $v$ is a real-valued function. Nevertheless, the assumption that space changes can be separated from changes in time is generally not realistic for earthquake activity.

In the nonparametric Bayesian framework, the estimation of $ \mu(x,y,t)$ can be done using the Generalized Spatial Dirichlet Process introduced by \cite{duan2007generalized}. Their approach allows modeling different probability density functions across space and discretized time, where the weights and point masses in the stick-breaking representation \citep{sethuraman1994constructive} are assumed to vary smoothly over space and time. However, their formulation requires modeling the weights through an auxiliary threshold scheme based on a collection of independent stationary Gaussian random fields. Since we keep the mass points fixed in our work, we have chosen to use the representation proposed by \cite{chakrabarti2024graphical}, which allows the weights to vary through a hierarchical model with beta distributions.

It is also worth mentioning the work by \cite{fonseca2017dynamic}, which deals with the spatio-temporal problem that occurs when point data are not directly observed and only aggregated counts are available. In this context, \cite{kottas2007bayesian} proposed a nonparametric Bayesian approach to estimate the time-homogeneous spatial Poisson-process intensity function using Dirichlet processes, work that has been extended to the inhomogeneous case by \cite{xiao2015modeling} using dependent Dirichlet processes (DDP) \citep{maceachern1999dependent}. If the reader is interested in DDP and their stick-breaking representation, we suggest consulting \cite{quintana2022dependent}.

The main difference between the methodology of \cite{xiao2015modeling} and the one proposed here lies in the DDP structure; while \cite{xiao2015modeling} keeps the mixture weights fixed over time and allows the component locations to vary, we instead keep the locations fixed and allow the weights to evolve over time, as discussed in Section \ref{sec:Methodology}.

The idea of keeping the mixture locations constant over time is reasonable, given that regions where tectonic plates interact require very long periods to undergo significant changes. Therefore, the main seismogenic regions will remain stable in their locations. When a change in seismicity occurs, for example, due to a strong earthquake or a slow slip event (SSE), the weight associated with the mixture component in these regions will deviate from zero, but only during the periods near the occurrence of the event. This, combined with the fact that the number of parameters to be estimated using the GDP is smaller, makes the approach presented in this work particularly attractive in the context of seismology.

The modeling of spatio-temporal point processes is not limited to seismology. In particular, recent works such as \cite{gilardi2024nonseparable,medialdea2025fully} have proposed models based on spatial Cox processes \citep{Diggle2013LGCP} to model ambulance interventions and wildfires. Using a frequentist approach, their proposals also allow the incorporation of exogenous variables and planar geometries.

Also, in the context of self-exciting point processes, \cite{briz2024point} proposed a purely temporal model that allows the self-exciting mechanism to be activated and deactivated over time, thereby reducing underestimation of the productivity parameter in a seismological context.

Finally, we mention the works of \cite{zhuang2019semiparametric,escudero2025crime}, which focused on modeling crime data, using log-Gaussian Cox and Hawkes processes with inhomogeneous separable spatio-temporal background intensities. This contrasts with the non-separable intensity proposed in this work based on a GDP, which highlights the novelty of our approach.

\section{Methodology}
\label{sec:Methodology}
As is pointed out by \cite{Fox2016Spatially}, the usual approach to model the mainshocks (i.e., earthquakes that are not triggered by other earthquakes, unlike aftershocks), is to use an inhomogeneous Poisson Process in space and constant over time with intensity $\mu(x,y)$, and the rate of aftershocks is modeled using a triggering function $\nu$ depending on the magnitude  ($M$) which decays according to the spatial and temporal distance between earthquakes. Then, the intensity function in the ETAS model is given by 
\begin{equation}
\label{eq:IntensityHawkes}
\lambda^*_{\mu,\nu}(t,x,y|H_t)=\mu(x,y)+\sum_{\{i:t_i<t\}} \nu(t-t_i,x-x_i,y-y_i,M_i), 
\end{equation}
where $(x_i,y_i)$ are the coordinates of the epicenter, $t_i$ the time of occurrence of the event $i$, with $i=1,2,...,N$, and $H_t$ denotes the history of earthquakes locations, times and magnitudes up to time $t$. Physically, $\mu(x,y)$ denotes the background seismicity, which refers to the earthquake rate that is unaffected by the occurrence of any other seismic event. The triggering function $\nu$ is described as the decay of an earthquake’s ability to generate aftershocks over time (i.e., it is based on the Omori Law \citep{utsu1995centenary}). Then, only earthquakes with magnitude exceeding a $M_u$ threshold are considered, based on the magnitude completeness provided by the Gutenberg-Richter Law\citep{gutenberg1949seismicity}. Then, the log-likelihood of the model is given by 
\begin{equation}
\label{eq:LogZhuang}
\ell(\mu,\nu):=\sum_{k=1}^n \log(\lambda^*_{\mu,\nu}(t_k,x_k,y_k|H_{t_k}))-\int_0^T \iint_D \lambda^*_{\mu,\nu} (t,x,y|H_t)dxdydt,
\end{equation}
where $(0,T)$ is the observation time interval and $D\subset \mathbb{R}^2$ denotes the observed region.

As we mentioned in Section \ref{sec:Intro}, the catalog from \cite{sawires2019updated} allows us to focus solely on the Poissonian intensity function. Therefore, in the rest of the text, we will work with a inhomogeneous Poisson Process (NHPP) with intensity function defined as: 
\begin{equation}
\label{eq:IntensityNHPP}
\lambda(t,x,y)=\mu(x,y,t),
\end{equation}
to model the background seismicity function.

It is important to note that if we have a sample $X=\{(x_i,y_i,t_i)\}$, where $(x_i,y_i)$ are observed in a region $D\subset \mathbb{R}^2$, with $i=1,2,...,N$ and in a time interval $(0,T)$ from a NHPP, the likelihood \citep{daley2003introduction} is given by  
\begin{equation}
\label{eq:loglike}
L(X|\mu) = e^{-\int_0^T \iint_D  \mu(x,y,t)dtdxdy}\prod_{i=1}^N \mu(x_i,y_i,t_i),    
\end{equation}
which, in general, is enormously computationally expensive because it requires solving a triple integral. Consequently, authors such as \cite{kottas2007bayesian,zhuang2002stochastic,ross2022semiparametric}, among others, have assumed, in the stationary case, that $\mu(x,y)$ is a bounded function. This form allows modeling it as 
$$\mu(x,y)=\gamma f(x,y),$$
where $\gamma$ is in $\mathbb{R}^+$, and $f(x,y)$ is a density function.  Additionally, if we assume that the domain of $f(x,y)$ is $D$, then the likelihood for the homogeneous case (in time) Equation \eqref{eq:loglike} can be rewritten as 
$$
L(X|\mu) = e^{-\gamma T} \gamma^N  \prod_{i=1}^N f(x_i,y_i),    
$$
which significantly reduces the computational cost, and then the idea of using the Dirichlet process to model $\mu$ as in \cite{kottas2007bayesian} or \cite{ross2022semiparametric} for a Bayesian approach emerges. 

As discussed in Section \ref{sec:Intro} following the idea of writing $\mu(x,y)=\gamma f(x,y)$, authors such as \cite{li2024multivariate} and \cite{xiao2015modeling} have proposed defining $\mu$ piecewise constant in time, i.e., if we have a partition of the interval (0,T) given by times $0=S_1,S_2,...,S_{p+1}=T$, then we can define $\mu$ as 
\begin{equation}
\label{eq:MuDefinition}
\mu(x,y,t)=\sum_{p=1}^P \gamma_p f_p(x,y) \mathbbm{1}(S_p<t<S_{p+1}),   
\end{equation}
where $\mathbbm{1}(\cdot)$ is the indicator function, which is the definition of $\mu$ that is used in our work. This definition aims to work with the inhomogeneities over space and time, but also allows us to rewrite Equation \eqref{eq:loglike} as 
\begin{equation}
\label{eq:Likelihood}
L(X|\mu) = \prod_{p=1}^P \Big( e^{-\gamma_p |I_p|} \gamma^{n_p}    \prod_{i=1}^{n_p} f_p(x_{p,i},y_{p,i})\Big),
\end{equation}
where $I_p=(S_p,S_{p+1})$, $|I_p|$ is the width of the partition, $X_p=\{(x_{p,i},y_{p,i})\}$ with $i=1,...,n_p$ is the subsample of $X$ of elements where $t_i$ is in $I_p$ for $p=1,...,P$.

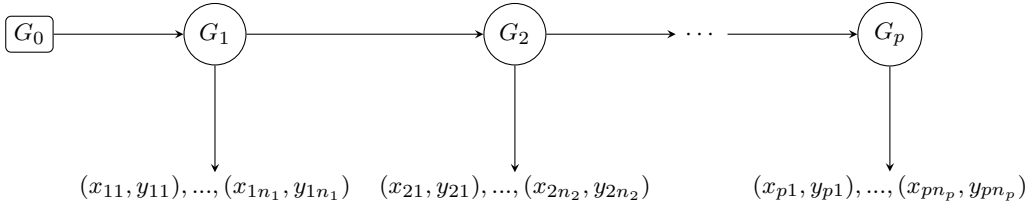
\begin{figure}
\centering

\begin{tikzpicture}[>=stealth, every node/.style={font=\small}]
  % matrix layout: first row = G nodes, second row = observed y's
  \matrix (m) [matrix of nodes,
    row sep=1.4cm,        % vertical distance between rows (increase if needed)
    column sep=0.3cm,     % horizontal distance between columns
    nodes={anchor=center},
  ]{
    \node[draw,rectangle,rounded corners=2pt,inner sep=3pt] (G0) {$G_0$}; &
    \node[draw,circle,minimum size=20pt] (G1) {$G_1$}; &
    \node[draw,circle,minimum size=20pt] (G2) {$G_2$}; &
    \node (dots) {$\cdots$}; &
    \node[draw,circle,minimum size=20pt] (Gp) {$G_p$}; \\
    % second row: aligned under each corresponding circle (use empty cell under G0)
    \node (empty) {}; &
    \node[inner sep=1pt, align=center] (y1) {$(x_{11},y_{11}),...,(x_{1n_1},y_{1n_1})$}; &
    \node[inner sep=1pt, align=center] (y2) {$(x_{21},y_{21}),...,(x_{2n_2},y_{2n_2})$}; &
    \node (phant) {}; &
    \node[inner sep=1pt, align=center] (yp) {$(x_{p1},y_{p1}),...,(x_{pn_p},y_{pn_p})$}; \\
  };

  % horizontal arrows between G's
  \draw[->] (G0) -- (G1);
  \draw[->] (G1) -- (G2);
  \draw[->] (G2) -- (dots);
  \draw[->] (dots) -- (Gp);

  % vertical arrows from selected G's to their observations
  \draw[->] (G1) -- (y1);
  \draw[->] (G2) -- (y2);
  \draw[->] (Gp) -- (yp);

  % optional: small styling tweaks (uncomment to use)
  % \node[below=2pt of y1] {}; % fine tune if needed
\end{tikzpicture}
\caption{Graphical Dirichlet Process denoting time dependency}
\label{fig:GPD_Graph}
\end{figure}

As can be seen in this definition of $\mu$, we have created non-exchangeable subsamples of $X$ that are dependent on $\prod_{i=1}^{n_p} f_p(x_{p,i},y_{p,i})$, which is a reasonable model if dependency is assumed. The temporal dependency used in this work can be described as a particular case of GDP, for which the directed acyclic graph is shown in Figure \ref{fig:GPD_Graph}. 

The GDP \citep{chakrabarti2024graphical} allows working with a more general DAG. Nevertheless, for our purposes, this version is adequate, and it also coincides with structures previously studied in the nested Chinese restaurant process \citep{blei2010nested} or the nested Hierarchical Dirichlet Processes \citep{paisley2014nested} that are encompassed in the GDP framework.

As mentioned by \cite{muller2015bayesian}, and since Dirichlet processes are discrete with probability one, it is awkward to estimate continuous densities. This limitation can be overcome using  DPM, which is generalized in \cite{chakrabarti2024graphical} to the GDP. The authors of \cite{chakrabarti2024graphical} proved that the structure presented in Figure \ref{fig:GPD_Graph} used to define a GDP mixture is 
\begin{align*}
    G_1 |\alpha_1,G_0 &\sim DP(\alpha_1,G_0),\\
    G_p |\alpha_p,G_{p-1}&\sim DP(\alpha_p,G_{p-1}),
\end{align*}
and for each group $p$ we have
\begin{align*}
\theta_{p,i}|G_p &\stackrel{\text{ind}}{\sim} G_p,\\
(x_{p,i},y_{p,i})|\theta_{p,i}&\sim F(\theta_{p,i}),    
\end{align*}
which, by the stick-breaking representation, can be recovered as the limit of the hierarchical finite mixture model presented in Equation \eqref{eq:Hierarchichal} when the number of $L$ mixture elements tends to infinity. Then, the addition of $\gamma_p$ variables, a priori independent of all $G_p$ processes in our hierarchical model, is given by
\begin{align}
\label{eq:Hierarchichal}
\alpha_1|\alpha_0&\sim \text{Gamma}(\alpha_0,1),\\
\nonumber
\beta_1|\alpha_1&\sim \text{Dir}\Big(\frac{\alpha_1}{L},...,\frac{\alpha_1}{L}\Big),\\
\nonumber
\alpha_{t}|\alpha_{t-1}&\sim \text{Gamma}(\alpha_{t-1},1),\\
\nonumber
\beta_p|\alpha_p,\beta_{p-1}&\sim \text{Dir}(\alpha_p (\beta_{p-1,1},\beta_{p-1,2},...,\beta_{p-1,L})),\\
\nonumber
\psi_l|G_0&\sim G_0,\\
\nonumber
z_{p,i}|\beta_p&\sim \text{Cat}(1:L,\beta_p),\\
\nonumber
\gamma_p&\sim \text{Gamma}(\gamma_0 k,k),\\
\nonumber
\{(x_{p,i},y_{p,i})|z_{p,i}, \{\psi_l\}_{l=1}^L, \gamma_p\}_{i=1}^{n_p}&\sim \mathcal{F}(\{z_{p,i}\}_{i=1}^{n_p}, \{\psi_l\}_{l=1}^L,\gamma_p) \text{ for all $p$ in \{1,2,..,P\}},
\end{align}
where $\{(x_{p,i},y_{p,i})|z_{p,i}, \{\psi_l\}_{l=1}^L, \gamma_p\}_{i=1}^{n_p}\sim \mathcal{F}(\{z_{p,i}\}_{i=1}^{n_p}, \{\psi_l\}_{l=1}^L,\gamma_p) $ has the following probability density function 
$$f((x_{p,1},y_{p,1}),...,(x_{p,n_p},y_{p,n_p}))=e^{-\gamma_p|I_p|} \gamma_t^{n_p}\prod_{i=1}^{n_p}\Big( \prod_{l=1}^L \phi_{\psi_l}(x_{p,i},y_{p,i})^{\mathbbm{1}(z_{pi}=l)}\Big),$$ 
and as we previously introduced in Equation \eqref{eq:Likelihood} we have
$$f(X|\boldsymbol \alpha,\boldsymbol \beta,\boldsymbol\gamma, \boldsymbol \psi, \mathbf{Z})=\prod_{p=1}^P \Big \{e^{-\gamma_p|I_p|} \gamma_t^{n_p}\prod_{i=1}^{n_p}\Big( \prod_{l=1}^L \phi_{\psi_l}(x_{p,i},y_{p,i})^{\mathbbm{1}(z_{pi}=l)}\Big)\Big\}.$$

In this work, we assume that the functions $\phi_{\mu_l,\Sigma^{-1}_l}$, $l=1,2,...,L$ are  Gaussian probability densities with mean $\mu_l$ and covariance matrix $\Sigma_l$. Also, to facilitate the notation, we write $\psi_l=(\mu_l,\Sigma_l^{-1})$ and $\phi_{\psi_l}$. Although this could break the assumption that $f$ has support $D$, we assume, as in \cite{ross2022semiparametric}, that the probability outside of $D$ is negligible and therefore Equation \eqref{eq:Likelihood} is still valid. 

The main reason for assuming that $f_{\psi_l}$ is Gaussian is to obtain the full conditional of $\psi_l$ using conjugate analysis, which will allow its straightforward incorporation into a Markov chain Monte Carlo (MCMC) scheme. 

It is worth mentioning that \cite{kottas2007bayesian} proposed a bivariate beta distribution to guarantee bounded support in $D$. Nevertheless, based on the previous paragraph, we decided to follow the approach proposed by \cite{ross2022semiparametric}.

For all our examples, we define $G_0\sim \text{NIW}(\mu_0,\eta,\Sigma_0,\nu)$ where NIW denotes a Normal Inverse Wishart distribution, we also denote this density by $f_{\text{NIW}}(\cdot)$. While the prior distribution reflects our uncertainty about the parameters before observing the data, this uncertainty is updated once the sample is observed through Bayes' theorem, as detailed in \cite{bernardo1994bayesian}, which defines the posterior distribution, i.e.,
$$f(\boldsymbol \alpha,\boldsymbol \beta,\boldsymbol\gamma, \boldsymbol \psi, \mathbf{Z}| X )=\frac{f(X|\boldsymbol \alpha,\boldsymbol \beta,\boldsymbol\gamma, \boldsymbol \psi, \mathbf{Z})f(\boldsymbol \alpha,\boldsymbol \beta,\boldsymbol\gamma, \boldsymbol \psi, \mathbf{Z} )}{f(X) }.$$
Since the denominator is only a normalizing constant and does not depend on the parameters of interest, the posterior distribution is fully characterized by the following expression

\begin{align}
\label{eq:Post}
f(\boldsymbol \alpha,\boldsymbol \beta,\boldsymbol\gamma, \boldsymbol \psi, \mathbf{Z}| X ) &\propto f(X|\boldsymbol \alpha,\boldsymbol \beta,\boldsymbol\gamma, \boldsymbol \psi, \mathbf{Z})f(\boldsymbol \alpha,\boldsymbol \beta,\boldsymbol\gamma, \boldsymbol \psi, \mathbf{Z} ) \\ \nonumber
&=\alpha_1^{\alpha_0-1}e^{-\alpha_1}
\frac{\displaystyle\prod_{l=1}^L\beta_{1l}^{\frac{\alpha_{1}}{L}-1}}{\text{B}(\frac{\alpha_{1}}{L},\frac{\alpha_{1}}{L},...,\frac{\alpha_{1}}{L})} 
\Big(\prod_{p=1}^{P-1} \frac{\alpha_{p+1}^{\alpha_{p}-1}e^{-\alpha_{p+1}}}{\Gamma(\alpha_{p})}\Big) \\ \nonumber
&\Big(\prod_{p=2}^P \frac{\displaystyle\prod_{l=1}^L\beta_{pl}^{\alpha_{p}\beta_{p-1,l}-1}}{\text{B}(\alpha_{p}\beta_{p-1,1},...,\alpha_{p}\beta_{p-1,L})}  \Big)
\Big(\prod_{l=1}^L f_{\text{NIW}}(\psi_l) \Big)\\ \nonumber
&\Big(\prod_{p=1}^P \prod_{i=1}^{n_p}\prod_{l=1}^L \beta_{pl}^{\mathbbm{1}(z_{pi=l})} \Big) \Big(\prod_{p=1}^P \gamma_p^{\gamma_0k-1}e^{-\gamma_pk} \Big)\\ \nonumber
& \prod_{p=1}^P e^{-\gamma_p|I_p|} \gamma_p^{n_p}\prod_{i=1}^{n_p}\Big( \prod_{l=1}^L \phi_{\psi_l}(x_{p,i},y_{p,i})^{\mathbbm{1}(z_{pi}=l)}\Big),
\end{align}
where 
\begin{align*}
\boldsymbol \alpha&=(\alpha_1,...,\alpha_P),\\
\boldsymbol \gamma&=(\gamma_1,...,\gamma_P),\\
\boldsymbol \psi&=(\psi_1,...,\psi_L),\\
\boldsymbol Z&=(z_{1,1},...,z_{1,n_1},z_{2,1},...,z_{2,n_2},...,z_{P,n_P}),\\
\boldsymbol \beta&=\begin{pmatrix},
\beta_{11}&\hdots&\beta_{1L}\\
\vdots&\ddots&\vdots\\
\beta_{P1}&\hdots&\beta_{PL}=
\end{pmatrix}=\begin{pmatrix}
\beta_{1}\\
\vdots\\
\beta_{P}
\end{pmatrix},
\end{align*}
To facilitate efficient sampling from the posterior distribution, we derive the full conditional distributions of the model parameters using the definition of conditional probability. In particular, the full conditional distribution of $\gamma_1$ is given by
$$f(\gamma_1|\alpha,\boldsymbol \beta,\boldsymbol\gamma_{-1}, \boldsymbol \psi, \mathbf{Z}, X )= \frac{f(\boldsymbol \alpha,\boldsymbol \beta,\boldsymbol\gamma, \boldsymbol \psi, \mathbf{Z}| X )}{f(\boldsymbol \alpha,\boldsymbol \beta,\boldsymbol\gamma_{-1}, \boldsymbol \psi, \mathbf{Z}, X )}\propto f(\boldsymbol \alpha,\boldsymbol \beta,\boldsymbol\gamma, \boldsymbol \psi, \mathbf{Z}| X ),$$
where $\boldsymbol{\gamma}_{-1}$ denotes the vector $\boldsymbol{\gamma}$ without its first component, i.e., the full conditional distribution of a parameter is proportional to the posterior distribution, treating the remaining parameters as constants. Therefore, it is easy to see that the full conditional for $\gamma_p$ for all $p=1,...,P$ is distributed as $\text{Gamma}(\gamma_0k+n_p,k+|I_p|)$.

For $\psi_l$ with $l=1,..,L$ the full conditional is given by
$$\pi(\psi_l|\cdot)\propto \prod_{p=1}^P \prod_{i=1}^{n_p}\Big( \prod_{l=1}^L \phi_{\psi_l}(x_{p,i},y_{p,i})^{\mathbbm{1}(z_{pi}=l)}\Big) f_{\text{NIW}}(\psi_l),  $$
which corresponds to independent and identically distributed (i.i.d.) Gaussian samples, which is a famous result of conjugated analysis that can be consulted in \cite{murphy2022probabilistic}, and is distributed as $\text{NIW}(\mu_l,\eta_l,\Sigma_l,\nu_l)$ where 
\begin{align*}
    \mu_l&=\frac{\eta\mu_0+m_l \widehat\mu_{xy}}{\eta+m_l},\\
    \eta_l&=\eta+m_l,\\
    \nu_n&=\nu+m_l,\\
    \Sigma_l&=\Sigma_0+S+\frac{\eta m_l}{\eta+m_l}(\widehat\mu_{xy}-\mu_0)(\widehat\mu_{xy}-\mu_0)^T,
\end{align*}
defined by 
\begin{align*}
m_l&=\displaystyle\sum_{p=1}^P\sum_{i=1}^{n_p} \mathbbm{1}(z_{p,i}=l),\\
\widehat\mu_{xy}&= \frac{\displaystyle\sum_{p=1}^P\sum_{i=1}^{n_p} (x_{p,i},y_{p,i})\mathbbm{1}(z_{p,i}=l) }{m_l}, \\
S&= \displaystyle\sum_{p=1}^P\sum_{i=1}^{n_p} ( (x_{p,i},y_{p,i}) -\widehat\mu_{xy})((x_{p,i},y_{p,i}) -\widehat\mu_{xy})^T\mathbbm{1}(z_{p,i}=l).
\end{align*}

It is important to note that the $\psi_l$ for $l=1,...,L$ are not independent as it can be seen from Equation \eqref{eq:Post} even when they were assumed a priori to be independent. Furthermore we want to highlight that their full conditional means take into account the samples $(x_{p,i},y_{p,i})$ for all $i=1,...,n_p$ and $p=1,2,...,P$, i.e. they are including information from the observed earthquakes among all the partitions.

In the case of $ \boldsymbol \alpha$ we have
\begin{align*}
    \pi(\alpha_1|\cdot)&\propto\frac{e^{-\alpha_1} \alpha_1^{\alpha_0 -1} \alpha_2^{\alpha_1}}{\Gamma(\alpha_1)}\frac{\displaystyle\prod_{l=1}^L\beta_{1l}^{\frac{\alpha_{1}}{L}}}{\text{B}(\frac{\alpha_{1}}{L},\frac{\alpha_{1}}{L},...,\frac{\alpha_{1}}{L})}, \\
    \pi(\alpha_P|\cdot)&\propto e^{-\alpha_P} \alpha_P^{\alpha_{P-1}-1 }\frac{\displaystyle\prod_{l=1}^L\beta_{Pl}^{\alpha_P\beta_{P-1,l}}}{\text{B}(\alpha_P\beta_{P-1,1},...,\alpha_P\beta_{P-1,L})} ,
\end{align*}
and for all $p$ in 2,...,$P-1$
\begin{align*}
    \pi(\alpha_p|\cdot)&\propto \frac{e^{-\alpha_p} \alpha_p^{\alpha_{p-1} -1 } \alpha_{p+1}^{\alpha_p}}{\Gamma(\alpha_p)}\frac{\displaystyle\prod_{l=1}^L\beta_{pl}^{\alpha_p\beta_{p-1,l}}}{\text{B}(\alpha_p\beta_{p-1,1},...,\alpha_p\beta_{p-1,L})}.
\end{align*}
Also, for $\boldsymbol \beta$ we have
$$\pi(\beta_1|\cdot)\propto \prod_{l=1}^L \beta_{1l}^{m_{1l}+\frac{\alpha_{1}}{L}-1} \frac{\displaystyle\prod_{l=1}^L\beta_{2l}^{\alpha_2\beta_{1,l}}}{\text{B}(\alpha_2\beta_{1,1},...,\alpha_2\beta_{1,L})}, $$
in a similar way, for $p=1,2,...,P-1$
$$\pi(\beta_{p}|\cdot)\propto \prod_{l=1}^L \beta_{pl}^{m_{pl}+\alpha_{p}\beta_{p-1,l}-1} \frac{\displaystyle\prod_{l=1}^L\beta_{p+1,l}^{\alpha_{p+1}\beta_{p,l}}}{\text{B}(\alpha_{p+1}\beta_{p,1},...,\alpha_{p+1}\beta_{p,L})}, $$
where \mbox{$m_{pl}=\displaystyle\sum_{i=1}^{n_p} \mathbbm{1}(z_{p,i}=l)$}. Meanwhile for $p=P$, since the term $p+1$ does not exist, we have $$\beta_P|\cdot \sim \text{Dirichlet}(m_{P1}+\alpha_P \beta_{P-1,1},...,m_{Pl}+\alpha_P \beta_{P-1,L}).$$

It is important to highlight that the full conditional distribution of $\beta_p$, for $p = 2, \dots, P-1$, depends on the neighboring vectors $\beta_{p-1}$ and $\beta_{p+1}$. Thus, even though an autoregressive structure is specified a priori, the resulting posterior dependence reinforces temporal coupling between adjacent coefficients. In combination with the values of $\boldsymbol{\alpha}$, this structure allows the model to capture smooth temporal behavior when $\boldsymbol{\alpha}$ is large, while permitting more abrupt changes in the density when $\boldsymbol{\alpha}$ is close to zero.

A limitation of this approach is that, as the norm of the time partition tends to zero, the number of observations within each interval becomes very small, which complicates estimation. However, this issue can be mitigated by imposing an informative prior on $\boldsymbol{\alpha}$, thereby enforcing a smoothness structure over the spatio-temporal evolution. This regularization may be particularly useful in seismic catalogs where large-magnitude slow slip events (SSEs) are observed without the occurrence of strong earthquakes. 

Finally,
$$\mathbb{P}(Z_{pi}=l|\cdot)\propto \beta_{pl} \phi_{\psi_l}(x_{p,i},y_{p,i}),$$
for $i=1,2,...,n_p$.

The autoregressive GDP of order 1 was selected for simplicity, however, as detailed in \cite{chakrabarti2024graphical} this framework can be extended to greater orders. Furthermore, as can be seen from the posterior distribution, the parameters are not independent even for non-contiguous time instants.

As discussed in \cite{chakrabarti2024graphical}, the full conditional distributions of $\alpha_i$, $i=1,...,P$, and $\beta_p$, for $p=1,...,P-1$ are not standard distributions and are the main bottleneck in the MCMC. Since all $\alpha_i$ are positive random variables, we propose a random walk step to explore them. According to \cite{chakrabarti2024graphical}, we used proposals based on the SALTSampler introduced by \cite{director2017efficient} to explore $\beta_p$. Then, we opted for a Hybrid MCMC \citep{robert1999monte} to sample the posterior distribution; the code is available at \hyperlink{github.com/isaiasmanuel/NHPP}{github.com/isaiasmanuel/NHPP}.

Since the discussion by Veen and Schoenberg \citep{veen2008estimation}, the use of hidden variables has increased in popularity to estimate the ETAS model. This is because, as they pointed out, the direct optimization of the likelihood, as originally proposed by Zhuang et al. \citep{zhuang2002stochastic}, is computationally expensive and numerically unstable.  The hidden variables included are 

\begin{align}
    \label{eq:chii}
    \chi_{ii}&=\begin{cases}
        1, &\text{if earthquake $i$ is a background event}\\
        0, &\text{otherwise}
    \end{cases},\\
    \label{eq:chiij}
    \chi_{ij}&=\begin{cases}
    1, &\text{if earthquake $i$ is an aftershock of $j$}\\
    0, &\text{otherwise}
    \end{cases}.
\end{align}
 These variables represent the branch structure of the earthquakes, and the log-likelihood of the ETAS model incorporating them is given by  

$$\ell_c \Big(\mu,\eta,\{\chi_{ii}\}_{i=1}^N,\{\chi_{ij}\}_{(i,j)\in I^2}\Big)=\ell^*_\text{O}(\mu,\{\chi_{ii}\}_{i=1}^N)+\ell^*_\text{I}(\nu,\{\chi_{ij}\}_{(i,j)\in I^2}\ ),$$
where $I^2$ denotes $\{1,2,...,N\}\times \{1,2,...,N\}$, where $\times$ denotes the Cartesian product, $\ell_\text{O}(\cdot)$ is the log-likelihood due to the offspring, and $\ell_\text{I}(\cdot)$ is the log-likelihood due to  immigrants. They are given by 

\begin{align}
    \label{eq:ellI}
    \ell^*_\text{I}(\mu,\{\chi_{ii}\}_{i=1}^N)&=  \sum_{i=1}^n \chi_{ii}\log(\mu(x_i,y_i))
    -\int_0^T \iint_S \mu(x,y,t) dxdydt \\
    \label{eq:ellO}
    \ell^*_\text{O}(\nu,\{\chi_{ij}\}_{(i,j)\in I^2}\ )&= \sum_{j=1}^n \Big[ \sum_{i>j}  \chi_{ij}  \log\Big(  \nu(t_i-t_j,x_i-x_j,y_i-y_j,M_j)     \Big) \\
    \nonumber
    &\hspace{1.5cm} - \int_{t_j}^T \iint_S  \nu(t-t_j,x-x_j,y-y_j,M_j) dxdydt   \Big].
\end{align}
Given this framework, \cite{veen2008estimation,li2020space} have used the EM algorithm for the ETAS model, and then \cite{molkenthin2022gp,ross2022semiparametric}  have introduced Bayesian hierarchical schemes where Gibbs proposals are used. 
It is important to highlight that only the Poissonian term of the model presented in this section can be easily incorporated into the ETAS framework by redefining $\ell^*_\text{I}$ as
$$ \ell^*_\text{I}(\mu,\{\chi_{ii}\}_{i=1}^N)= \sum_{p=1}^P \Big(\sum_{i=1}^n \chi_{ii}\log(\mu^p(x_i,y_i,t_i))
    -\int_0^T \iint_S \mu^p(x,y,t) dxdydt \Big).$$
However, in our case, this evaluation is not necessary to estimate the background intensity function because the features of the \cite{sawires2019updated} catalog allow us to assume simply that all the $\chi_{ij}$ are 0 for all $i \neq j$, whereas $\chi_{ii} = 1$ for all $i$.

\section{Numerical Experiments}

\subsection{Simulated Data}
\label{sec:Simulated}

To verify the performance of our model, we apply it to a synthetic earthquake catalog, with $D=(-5,10)\times(-5,10)$ and $t\in(0,10)$ with the following intensity function 
\begin{equation}
\label{eq:lambdasim}
\lambda(x,y,t)= \Big(50\mathbbm{1}(t<5)+100 \mathbbm{1}(t \geq 5) \Big) \Big( h(t)g_1(x,y)+ (1-h(t) ) g_2(x,y) \Big),    
\end{equation}
where $\mathbbm{1}(\cdot)$ denotes the indicator function and 

\begin{align*}
h(t)&=\frac{1}{1+e^{-\frac{t-T}{2}}},\\
g_1(x,y)&= \frac{2}{3}\phi_{(0,0), \mathbf{1}_{2\times 2}}(x,y)+\frac{1}{3}\phi_{(2,2), \mathbf{1}_{2\times 2}}(x,y),   \\
g_2(x,y)&= \frac{2}{3}\phi_{(6,2), \mathbf{1}_{2\times 2}}(x,y)+\frac{1}{3}\phi_{(4,6), \mathbf{1}_{2\times 2}}(x,y).   \\
\end{align*}
The idea behind the proposed $\lambda$ function is to construct a spatio-temporal intensity function capable of capturing the typical challenges observed in an earthquake catalog. For example, in \cite{banales2025inhomogeneous}, it has been shown that SSEs can modify seismicity. In particular, in southern Mexico, SSEs can last longer than one year, which motivates the function $h$ in Equation \eqref{eq:lambdasim}, allowing for a smooth transition between modes. Additionally, the number of recorded earthquakes may increase due to changes in instrumentation or aseismic transients, as discussed in Section \ref{sec:Intro}. In Mexico, improvements in instrumentation, as described in \cite{vergnolle2010slow,cruz2018seismogeodetic}, have increased the number of recorded earthquakes. This non-smooth change is reflected in Equation \eqref{eq:lambdasim} by changing the productivity parameter from 50 to 100 at time 5.

In this example we take for the Normal-inverse-Wishart distribution $\mu_0=(1,1)$, where the vector (1,1) was chosen arbitrarily to be inside $D$, $\Sigma_0=\mathbb{I}_{2}$, $\nu=3$ and $\eta=0.1$, where $\mathbb{I}_{2}$ is the $2\times2$ identity matrix, the idea is to have a non informative prior distribution for the parameters in the mixture. For all the $\gamma_p$ we take $\gamma_0=70$ and $k=0.1$, which results in a vague prior distribution and $\alpha_0=1$. Also, a regular partition of the interval $(0,T)$ was used with $P=8$ and $L=8$.  

\begin{figure}[!ht]
    \centering
    \includegraphics[width=1\linewidth]{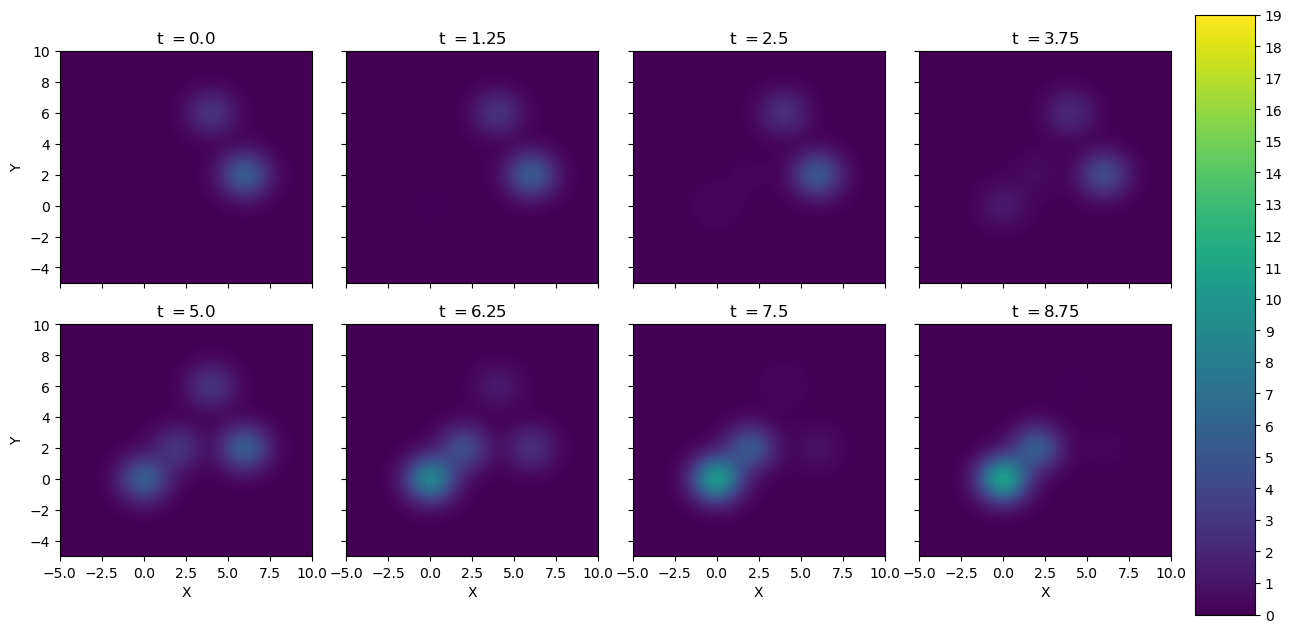}
    \caption{$\lambda$ evaluated at different times}
    \label{fig:lambdaMesh}
\end{figure}

Figure \ref{fig:lambdaMesh} shows the intensity function in the whole space at different times. This visual representation facilitates the observation of spatial and temporal changes. The stochastic simulation of this NHPP was generated using the thinning algorithm \citep{lewis1979simulation2, illian2008statistical}, which is summarized as follows
\begin{enumerate}
    \item Define $\lambda^*$ as the maximum of $\lambda(\cdot),$ i.e. $$\lambda^*:=\max_{(x,y)\in D, t\in (0,T)} \lambda(x,y,t).$$
    \item Simulate a sample $\{(x_i^s,y_i^s,t_i^s)\}_{i=1}^{N^*}$
    that follows a homogeneous Poisson Process (HPP) in $D$ over $(0,T)$ with intensity $\lambda^*$. For a deeper discussion about how to simulate an HPP consult \cite{bremaud2020point}.
    \item For each $i$ in $\{1,2,...,N^*\}$, the sample $(x_i^s,y_i^s,t_i^s)$ is accepted as a valid simulation of the NHPP with intensity function $\lambda(\cdot)$ with probability
    $$p_i=\frac{\lambda(x_i^s,y_i^s,t_i^s)}{\lambda^*}.$$
\end{enumerate}

The simulated catalog, consisting of 727 events, and the code to generate it can be consulted in 
\hyperlink{github.com/isaiasmanuel/NHPP}{github.com/isaiasmanuel/NHPP}.

\begin{figure}[!ht]
    \centering
    \includegraphics[width=1\linewidth]{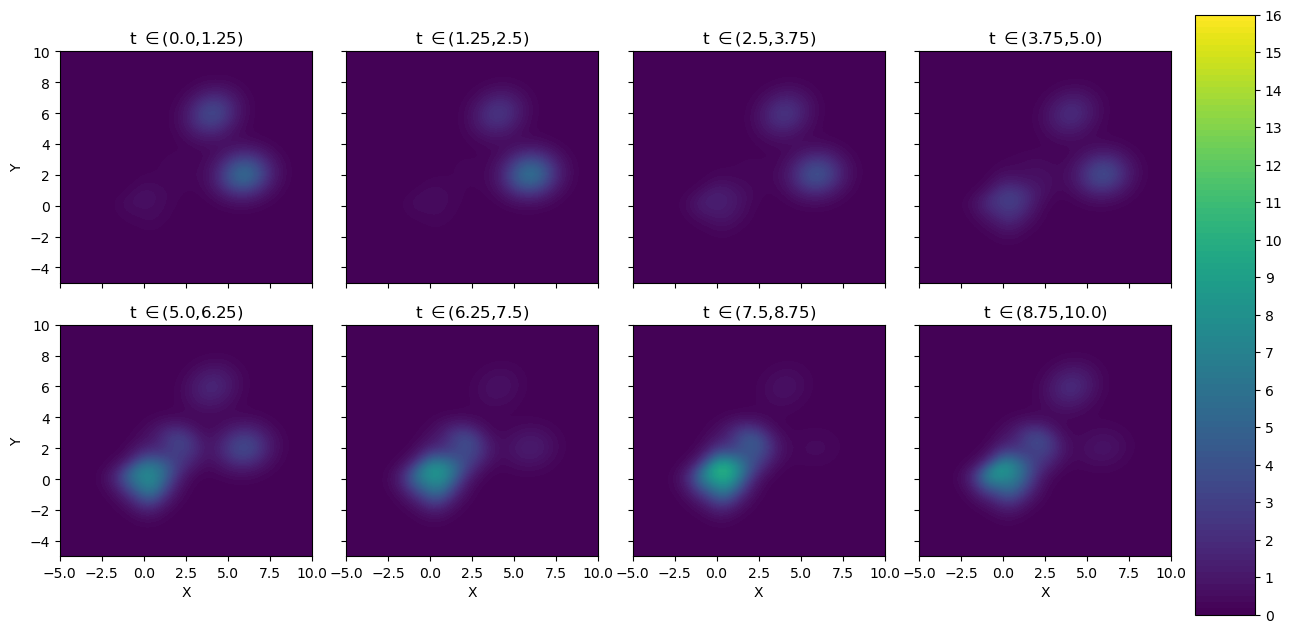}
    \caption{ Posterior mean of $\lambda$}
    \label{fig:lambdaMeshEstimaded}
\end{figure}

\begin{figure}[!ht]
    \centering
    \includegraphics[width=1\linewidth]{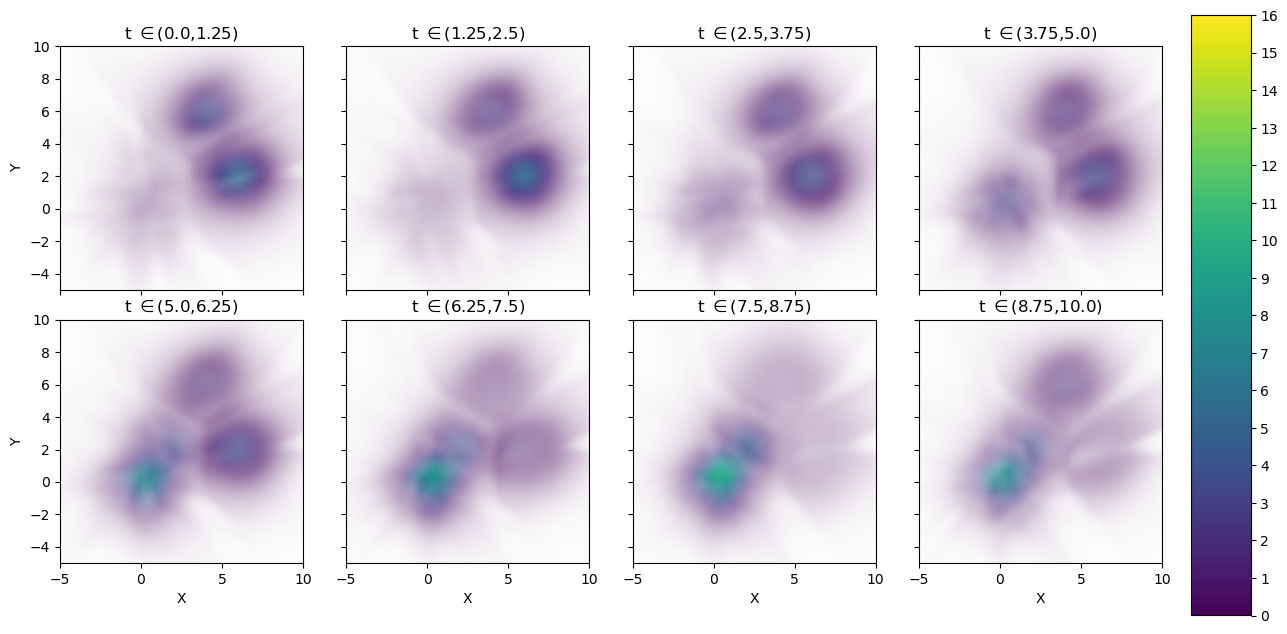}
    \caption{ Posterior mean of $\lambda$ with transparency based on the coefficient of variation using simulated data}
    \label{fig:lambdaMeshCV}
\end{figure}

Figure \ref{fig:lambdaMeshEstimaded} presents the posterior mean of $\gamma_pf_p$, as defined in Equation \eqref{eq:MuDefinition}. Then, to utilize the information in the posterior distribution more effectively, we present Figure \ref{fig:lambdaMeshCV}. The colors of the figure are the same as in Figure \ref{fig:lambdaMeshEstimaded}, but now we include a transparency based on the posterior coefficient of variation (CV), which is defined by
$$\text{CV}_p(x,y):= \frac{\sigma(x,y)}{\mu(x,y)}, $$
where $\mu$ and $\sigma$ are the mean and posterior variance of $\lambda(x,y,t)$ where $t\in (S_{p},S_{p+1})$. Then the transparency for each figure $p$ is given by
$$1-\frac{\text{CV}_p-\min_{x,y,p}\text{CV}_p}{\text{CV}_p}, $$
In other words, if the posterior mean is high compared to the posterior standard deviation, the color is saturated. The transparency increases as the average decreases relative to the standard deviation. 

\begin{figure}[!ht]
    \centering
    \includegraphics[width=1\linewidth]{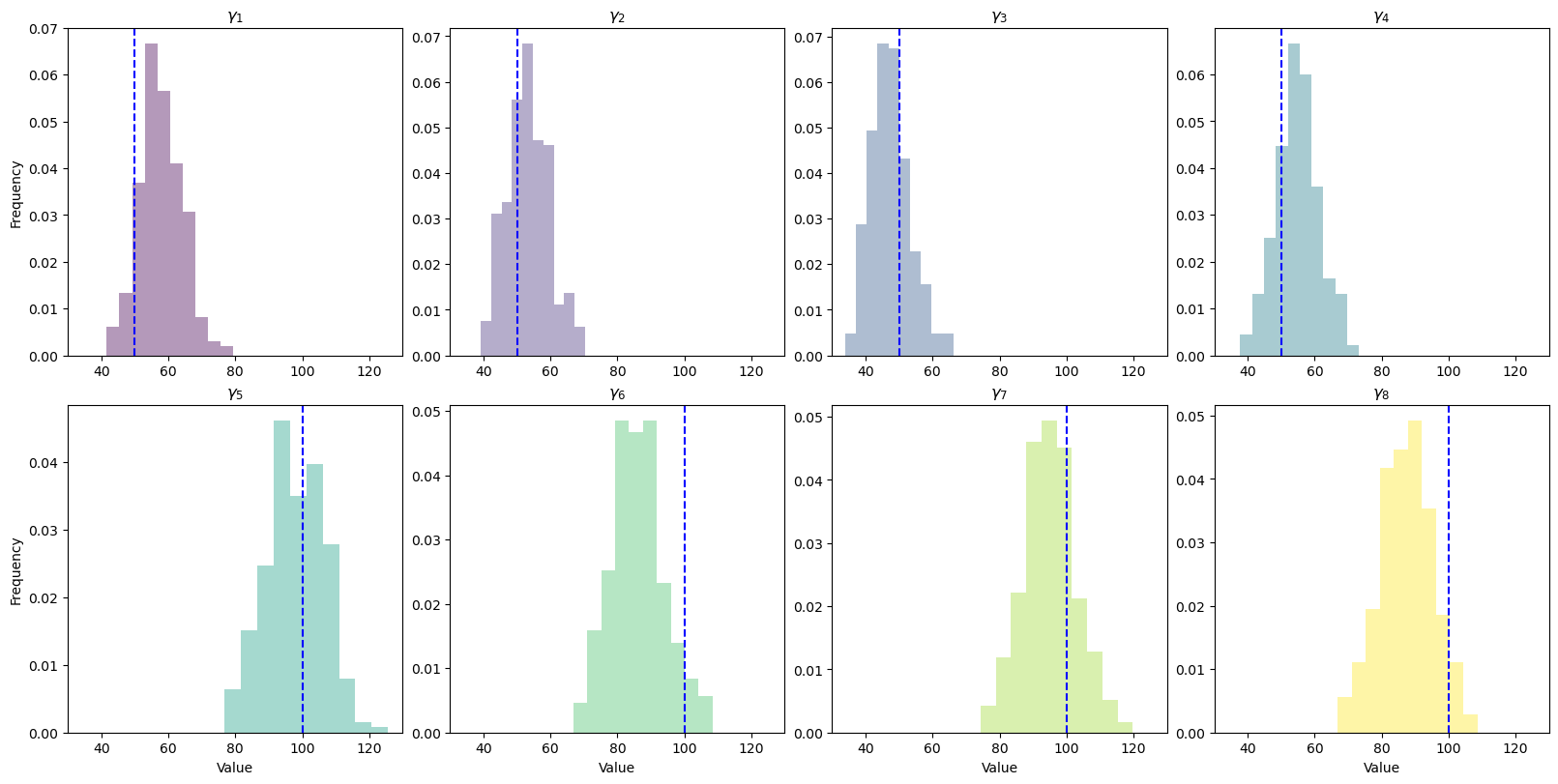}
    \caption{ Histogram of the posterior $\gamma_p$, $p=1,2,...,P$ using simulated data, blue dashed lines are the real $\gamma_p$ for each interval}
    \label{fig:gamma}
\end{figure}

As can be seen in Figure \ref{fig:lambdaMeshCV}, we have lower coefficient of variation values in regions with higher values of $\lambda$ (i.e., as expected, the regions with the highest number of points are better estimated). To complement Figure \ref{fig:lambdaMeshCV}, we also present Figure \ref{fig:gamma}, where the histograms of the marginal posterior samples are presented for each $\gamma_p$ and the true value $\gamma_p$ (vertical dashed blue line), according to Equation \eqref{eq:lambdasim}. Figure \ref{fig:gamma} reveals that the posterior distributions of $\gamma_p$ adequately contain the true value for all $p=1,2,...,P$.

%%%%%%%

\begin{figure}[ht]
\centering
\includegraphics[width=0.6\textwidth]{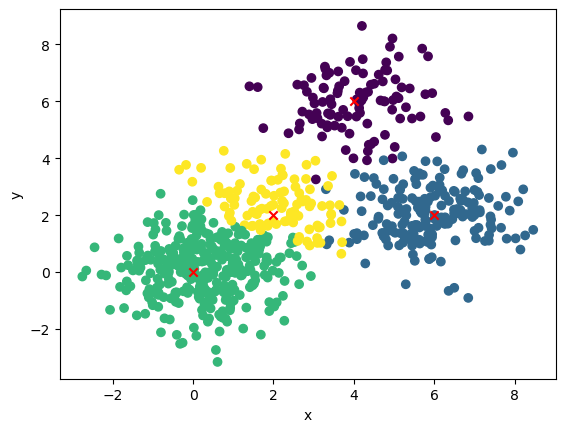}
\caption{4-groups clustering of the simulated data}
\label{fig:Cluster}
\end{figure}

In this work, we do not focus on the inference related to each element in the mixture. Instead, we are interested in inferring the intensity function $\lambda$. Nevertheless, it is worth mentioning that we observed the label switching problem, as expected in a finite mixture problem \citep{green1995reversible,zhang2004learning,bouguila2012fully}. For this reason, we have implemented the algorithm proposed by \cite{papastamoulis2010artificial} (the code is available at \hyperlink{github.com/isaiasmanuel/NHPP}{github.com/isaiasmanuel/NHPP}), and we also relabel by lexicographic order of the cluster mean in order to select the thinning in the chain. To read the discussion of the burn-in and thinning of the chain, please see the supplementary material.

 To validate our model, we present in Figure \ref{fig:Cluster} the clustering of the observed points. We show the mean of the Gaussian distributions (red crosses) used to define $\lambda$ in Equation \eqref{eq:lambdasim}. This clustering was obtained using the variation of information metric (VI) introduced by \cite{meilua2007comparing} and the library \textit{mcclust.ext} presented in \cite{wade2018bayesian}. As can be seen, we recover the four mixture components adequately, even when the mixture weights change over time.

We also performed clustering using Binder's loss, as proposed in \cite{binder1978bayesian,dahl2006model}. Nevertheless, we observed a substantial overestimation of the number of clusters, as it can be consoulted in the supplementary material, which is consistent with the discussions in \cite{wade2018bayesian,dahl2022search}. For this reason, we prefer to use VI. Finally, we would like to mention the papers \cite{socher2011spectral,duan2024spectral}, where new priors for clusters are proposed based on distance measures and Bayesian spanning forests that rely on spectral clustering, which may be of interest to readers primarily focused on clustering analysis.

From this example, we have seen how GDP can be used to accurately estimate the intensity function of an inhomogeneous spatial Poisson process. That is why, in the next section, we apply this same methodology to the \cite{sawires2019updated} catalog to recover the background seismicity function.

\subsection{Southern Mexico} As we discussed previously in Section \ref{sec:Intro}, different authors have pointed out that the background seismicity function could change over time by aseismic transients  \citep{matthews1988statistical,marsan2013slow,reverso2015detection,nishikawa2023development} or by strong earthquakes \citep{melgar2018deep, cruz2021short}; nevertheless, a common assumption in the ETAS model is to assume a constant background seismicity function  \citep{zhuang2002stochastic,ross2022semiparametric,molkenthin2022gp}. If this assumption is suitable for the seismicity in southern Mexico, we expect to see homogeneity between all $\gamma_pf_p$ for all $p=1,2,...,P$. Alternatively, if the assumption is not appropriate, we expect to observe changes over space or time related to the regions where we observed SSE or Strong earthquakes.

As reported by \cite{radiguet2012slow,radiguet2016triggering,cruz2021short}, the SSE in Guerrero, Mexico, between 2000/01/01 and 2015/12/31, had an approximate periodicity of 4 years; for this reason, in the GDP inference for the Mexican data, we use a regular time partition with $P=4$, thus obtaining an SSE at each time interval.

\begin{figure}
    \centering
    \includegraphics[width=0.8\linewidth]{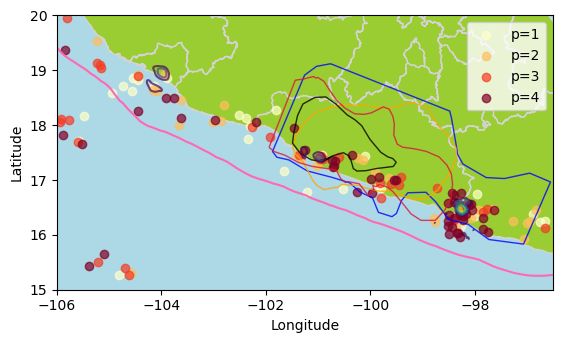}
    \caption{Epicenters from the \cite{sawires2019updated} catalog, SSE contours from \cite{radiguet2012slow,radiguet2016triggering}, slip distribution contours for earthquakes with $M\geq 7$ from \cite{Geological}, and the Middle America Trench using data from \cite{GEBCO}}
    \label{fig:Seismicity}
\end{figure}

The epicenters of pure strike-slip faulting and pure reverse faulting taken from \cite{sawires2019updated} are presented in Figure \ref{fig:Seismicity} (colored by their time interval $p$). The Middle America Trench \citep{GEBCO} is presented in fuchsia. Furthermore, the slip contour curves digitized from \cite{radiguet2012slow} of the 2001-2002, 2006, and 2009-2010 SSE are presented in blue, orange, and red, respectively (each 4 cm). In addition, the  15 cm slip contour of the 2014 SSE is presented in black, which was digitized from \cite{radiguet2016triggering}. Additionally, we present the slip distribution, with 1-meter resolution, for the earthquakes occurred in $2003/01/22$, $2012/03/20$, and $2014/04/18$, with magnitudes  7.6, 7.4 and 7.2, and epicenters  -104.1040E$^\degree$ 18.770N$^\degree$,  -98.2310E$^\degree$ 16.493N$^\degree$, -100.9723E$^\degree$ 17.397N$^\degree$, respectively, using data from \cite{Geological}.

The hyperparameters in the Normal Inverse Wishart were taken as $\nu=3$, $\eta=0.01$, $\Sigma_0=\mathbb{I}_2$, $\mu_0=(-102,17)$, which was arbitrarily chosen to be a point near the center of the study region along the trench, and we restrict $\mu$ to be in the observed domain (i.e., $(-105.5\text{E}^\degree,-96.5\text{N}^\degree )\times (15\text{E}^\degree,19.5\text{N}^\degree )$). 

We take $\alpha_0=1$ as in the previous example. For $\gamma_p$, we follow \cite{ross2022semiparametric}, and we take $\gamma_0=1$ and $k=0.01$. Finally, we set $L=12$, similar to \cite{chakrabarti2024graphical}, where $L=10$ was used for a simulation example with a sample size comparable to ours was estimated. However, we ran MCMC with a larger $L$ value, but we did not observe improvements in the estimated intensities. In contrast, the MCMC worsened due to the label switching, as there were constantly mixture components with 0 observations allocated.

\begin{figure}
    \centering
    \includegraphics[width=1\linewidth]{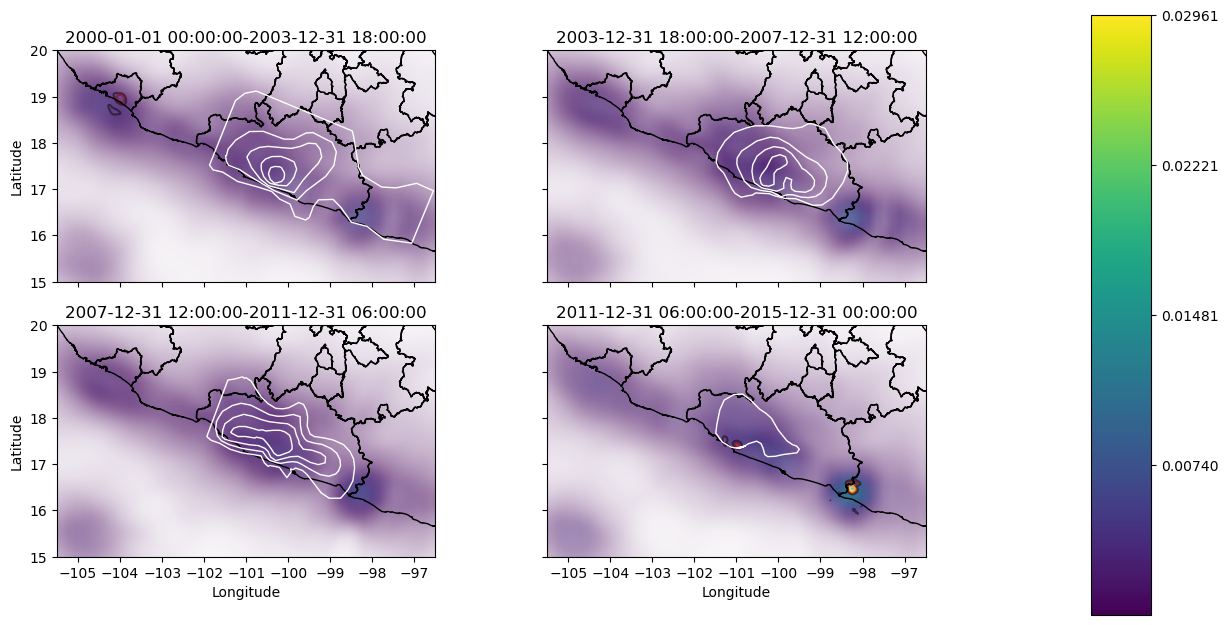}
    \caption{Posterior mean of $\lambda$ in $\frac{\text{events}}{\text{days deg$^2$}}$ with transparency based on the coefficient of variation using the catalog of \cite{sawires2019updated}; the gray curves in $p=1,2,3$ are the slip contours at 4 cm intervals for the corresponding SSE in Figure \ref{fig:Seismicity} and the curve in $p=4$ is the slip contour at 15 cm}
    \label{fig:SeismicitySawires}
\end{figure}

Figure \ref{fig:SeismicitySawires} shows the posterior mean of $\gamma_p f_p$ for each $p=1,2,3,4$. It can be observed that the intensity function reaches its maximum near the slip distribution contours of the strong earthquakes in the interval $p=1$ and $p=4$. Furthermore, we can also appreciate that near the border of the SSE slips, the background seismicity function exhibits significant activity. Previously, the authors of \cite{fukuda2018variability} observed a similar behavior in the Boso Peninsula, Japan. The increase in seismic activity at the border of the SSE could be explained by stress redistribution caused by the largest SSE slips, which may reduce stress in some regions while increasing it in others, potentially triggering earthquakes where stress accumulated but was not released, as previously hypothesized in the region by \cite{cruz2021short,cruz2025seafloor,villafuerte2025slow}, among others.

\begin{figure}
    \centering
    \includegraphics[width=1\linewidth]{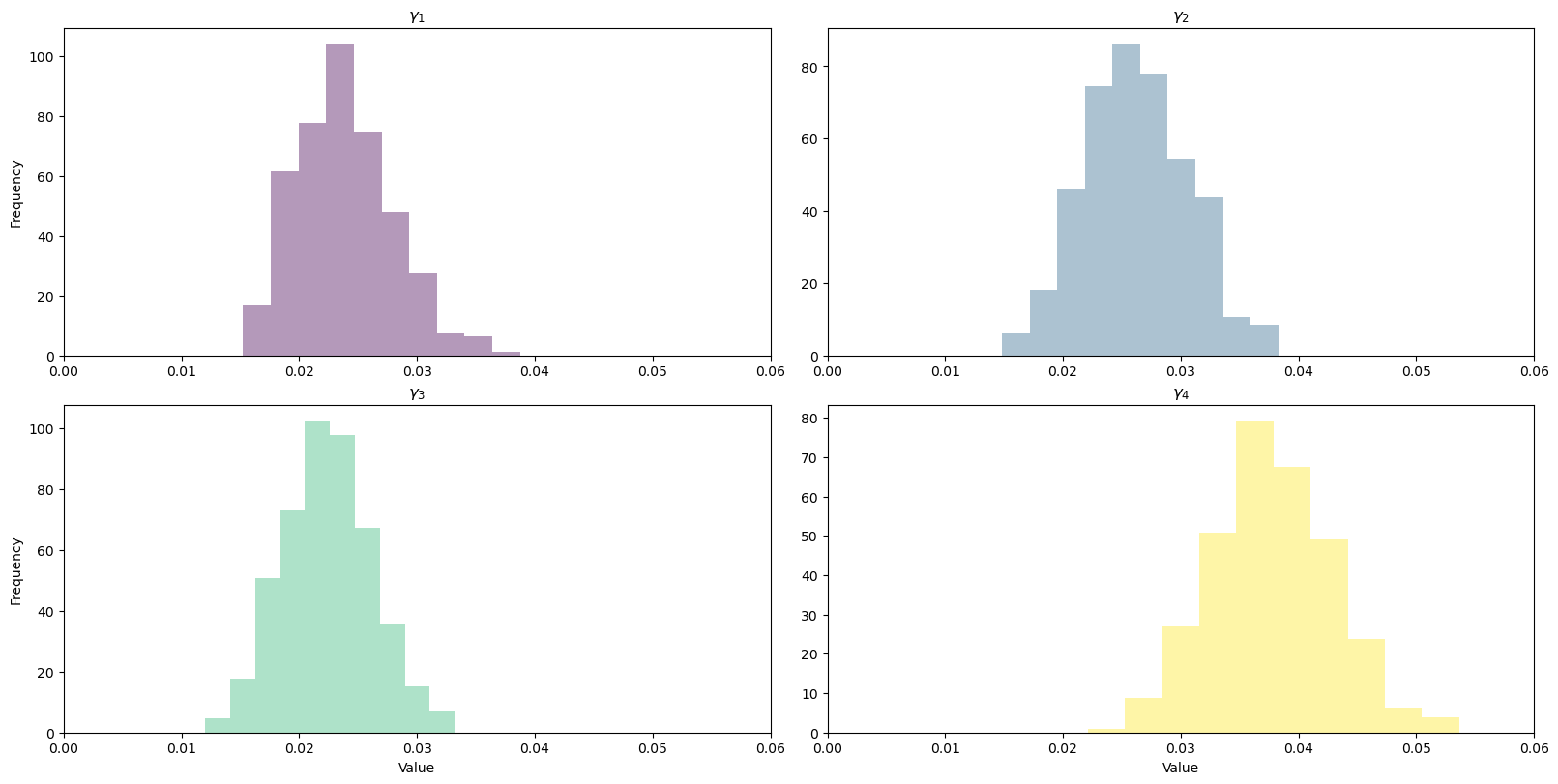}
     \caption{ Histogram of the posterior $\gamma_p$, $p=1,2,...,P$ using the catalog of \cite{sawires2019updated}}
    \label{fig:gammaSawires}
\end{figure}

While Figure \ref{fig:SeismicitySawires} shows the spatio-temporal changes, Figure \ref{fig:gammaSawires} presents the marginal posterior distributions of $\gamma_p$ for $p = 1, 2, \ldots, 4$. In particular, we observe a significant shift toward higher values in the $\gamma_4$ distribution, indicating an increase in the expected number of earthquakes during the last period, which is consistent with the occurrence of the two strong earthquakes observed in this interval.

\begin{figure}[ht]
\centering
\includegraphics[width=0.8\textwidth]{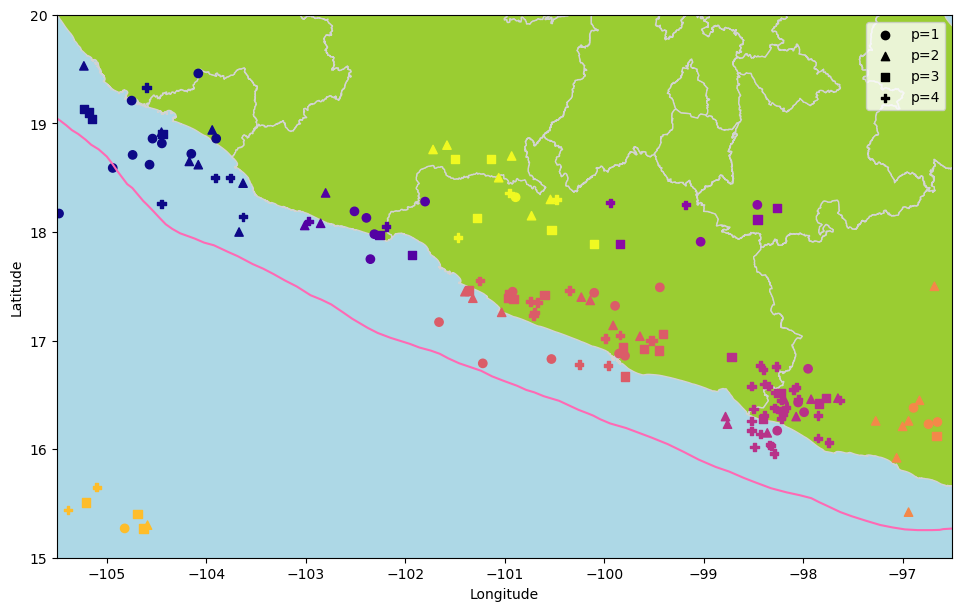}
\caption{Clustering into 8 groups using the catalog of \cite{sawires2019updated}, each color represents one group, and the geometric shapes represent the time period}
\label{fig:ClusterSawires}
\end{figure}

As we presented in Section \ref{sec:Simulated}, one of the main advantages of using a Dirichlet process mixture is that we are also clustering the events, as it is discussed in \cite{muller2015bayesian}, estimating $f_p$. In Figure \ref{fig:ClusterSawires}, we present the data from \cite{sawires2019updated} clustered in 8 groups using the VI metric as in the section \ref{sec:Simulated}.

The clusters may be correlated with seismogenic features reported in the literature. For example, earthquakes assigned to the magenta cluster in Figure \ref{fig:ClusterSawires} spatially correspond to the high b-value (1.50$\pm$0.10) region reported by \cite{legrand2021influence}. Specifically, earthquakes in the indigo cluster may be consistent with the events associated with the ultra-slow (USL) velocity layer reported by \cite{song2009subducting}. Additionally, the blue cluster corresponds spatially to seismicity in the Rivera plate subduction region.

Finally, the orange clusters correspond to the earthquakes in the catalog that occurred in the oceanic crust of the Cocos plate. The retrieval of this cluster was theoretically anticipated due to its geographical remoteness from the other clusters. However, the divergence from their interpretation is evident, as the orange cluster’s emergence is the primary focus, whereas for the rest of the clusters, their coincidence with the subducted plate’s shape and the seismogenic zones is the salient feature to emphasize.

%
%Finally, the pink clusters correspond to the earthquakes in the catalog that occurred in the oceanic crust of the Cocos plate. The retrieval of this cluster was theoretically anticipated due to its geographical remoteness from the other clusters.
%}

The code to replicate this section is available at \hyperlink{github.com/isaiasmanuel/NHPP}{github.com/isaiasmanuel/NHPP}, and for the discussion of the burn-in and thinning of the chain, please see the supplementary material.

\section{Conclusions}

Although the use of DPM for nonparametric Bayesian estimation of the background seismic activity function had been previously studied by \cite{ross2022semiparametric}, they only considered homogeneity in the time case. In this work, we have succeeded in considering temporal changes in the background activity function by including GDP \cite{chakrabarti2024graphical}. 

Furthermore, in the context of estimating the background seismic activity function with GDP, it has easily allowed us to identify seismogenic regions by only taking seismicity into account, which had not been previously discussed by \cite{ross2022semiparametric} and which is not intuitive to obtain following the approach presented in \cite{molkenthin2022gp}.

As discussed previously by \cite{molkenthin2022gp}, it is important to note that the Gaussian process approach requires the correlation structure to be specified, whereas the DP approach requires the number of mixture components to be determined. The use of a mixture of DPs can estimate a more flexible dependence structure, mainly due to local behaviors, than Gaussian processes. 

Our work can be viewed as an extension of the ideas presented in \cite{matthews1988statistical}, which proposed using an inhomogeneous Poisson process for the background seismicity function. The intensities estimated using the catalog by \cite{sawires2019updated} for southern Mexico have exhibited spatial and temporal changes associated with strong regular earthquakes and SSEs. Therefore, we can conclude that the assumption of a homogeneous Poisson process is not realistic for southern Mexico.

In future work, we will investigate the self-exciting dynamics of earthquakes. While the present study focuses on inhomogeneities in the background intensity function, studies such as \cite{ueda2021spatial,briz2025self} have demonstrated that spatially or spatio-temporally varying self-excitation parameters can improve the modeling of seismicity. Additionally, it is interesting to study different DDP structures as higher-order autoregressive GDP or consider the time arrival structure introduced in \cite{griffin2006order}.

\bibliography{sample.bib}       % Bibliography file 

\end{document}